\documentclass{emulateapj}
\usepackage{apjfonts}
\usepackage{color}

\newcommand {\mchi} {m_{\rm 3.6 \mu m} } 
\newcommand {\mchii} {m_{\rm 4.5 \mu m} }

\shorttitle{Faint 1.1 mm Sources}
\shortauthors{Wang et al.}

\begin{document}

\title{SXDF-ALMA 2-arcmin$^2$ Deep Survey: Stacking of Rest-Frame Near-Infrared Selected Objects}

\author{Wei-Hao Wang\altaffilmark{1,2}, Kotaro Kohno\altaffilmark{3,4}, Bunyo Hatsukade\altaffilmark{5,\dag}, Hideki Umehata\altaffilmark{3,6},
Itziar Aretxaga\altaffilmark{7}, David Hughes\altaffilmark{7}, Karina I.\ Caputi\altaffilmark{8}, James S. Dunlop\altaffilmark{9}, Soh Ikarashi\altaffilmark{8},
Daisuke Iono\altaffilmark{5,10}, Rob J.\ Ivison\altaffilmark{6,9}, Minju Lee\altaffilmark{11,5}, Ryu Makiya\altaffilmark{12,13}, Yuichi Matsuda\altaffilmark{5,10}, 
Kentaro Motohara\altaffilmark{3}, Kouichiro Nakanish\altaffilmark{5,10}, Kouji Ohta\altaffilmark{14}, Ken-ichi Tadaki\altaffilmark{15}, Yoichi Tamura\altaffilmark{3}, 
Tadayuki Kodama\altaffilmark{5,10}, Wiphu Rujopakarn\altaffilmark{12,16}, 
Grant W. Wilson\altaffilmark{17}, Yuki Yamaguchi\altaffilmark{3}, Min S.\ Yun\altaffilmark{17}, Jean Coupon\altaffilmark{18}, Bau-Ching Hsieh\altaffilmark{1}, and S\'ebastien Foucaud\altaffilmark{19}}

\altaffiltext{1}{Academia Sinica Institute of Astronomy and Astrophysics (ASIAA), No.\ 1, Sec.\ 4, Roosevelt Rd., Taipei 10617, Taiwan}
\altaffiltext{2}{Canada-France-Hawaii Telescope (CFHT), 65-1238 Mamalahoa Hwy, Kamuela, HI 96743, USA}
\altaffiltext{3}{Institute of Astronomy, University of Tokyo, 2-21-1 Osawa, Mitaka, Tokyo 181-0015, Japan}
\altaffiltext{4}{Research Center for the Early Universe, The University of Tokyo, 7-3-1 Hongo, Bunkyo, Tokyo 113-0033, Japan}
\altaffiltext{5}{National Astronomical Observatory of Japan (NAOJ), 2-21-1 Osawa, Mitaka, Tokyo 181-8588, Japan}
\altaffiltext{6}{European Southern Observatory, Karl-Schwarzschild-Str. 2, D-85748 Garching, Germany}
\altaffiltext{7}{Instituto Nacional de Astrof\'{i}sica, \'{O}ptica y Electr\'{o}nica (INAOE), Luis Enrique Erro 1, Sta.\ Ma.\ Tonantzintla, Puebla, Mexico}
\altaffiltext{8}{Kapteyn Astronomical Institute, University of Groningen, P.\ O.\ Box 800, 9700AV Groningen, The Netherlands}
\altaffiltext{9}{Institute for Astronomy, University of Edinburgh, Royal Observatory, Edinburgh EH9 3HJ UK}
\altaffiltext{10}{SOKENDAI (The Graduate University for Advanced Studies), 2-21-1 Osawa, Mitaka, Tokyo 181-8588, Japan}
\altaffiltext{11}{Department of Astronomy, The University of Tokyo, 7-3-1 Hongo, Bunkyo-ku, Tokyo 133-0033, Japan}
\altaffiltext{12}{Kavli Institute for the Physics and Mathematics of the Universe, Todai Institutes for Advanced Study, the University of Tokyo, Kashiwa, Japan 277-8583 (Kavli IPMU, WPI)}
\altaffiltext{13}{Max-Planck-Institut f\"{u}r  Astrophysik (MPA), Karl-Schwarzschild Str. 1, D-85741 Garching, Germany}
\altaffiltext{14}{Department of Astronomy, Kyoto University, Kyoto 606-8502, Japan}
\altaffiltext{15}{Max-Planck-Institut f\"{u}r extraterrestrische Physik (MPE), Giessenbachstrasse, D-85748 Garching, Germany}
\altaffiltext{16}{Department of Physics, Faculty of Science, Chulalongkorn University, 254 Phayathai Road, Pathumwan, Bangkok 10330, Thailand}
\altaffiltext{17}{Department of astronomy, University of Massachusetts, Amherst, MA 01003, USA}
\altaffiltext{18}{Astronomical Observatory of the University of Geneva, ch. d'Ecogia 16, CH-1290 Versoix, Switzerland}
\altaffiltext{19}{Center for Astronomy \& Astrophysics, Department of Physics \& Astronomy, Shanghai JiaoTong University, 800 Dongchuan Road, Shanghai 200240, China}
\altaffiltext{\dag}{NAOJ Fellow}

\begin{abstract}
We present stacking analyses on our ALMA deep 1.1 mm imaging in the SXDF using 1.6 $\mu$m and 3.6 $\mu$m 
selected galaxies in the CANDELS WFC3 catalog.   We detect a stacked flux of 
$\sim 0.03$--0.05 mJy,  corresponding to $L_{\rm IR}<10^{11}~L_\sun$
and a star formation rate (SFR) of $\sim15~M_\sun$ yr$^{-1}$ at $z=2$.  We find that galaxies brighter in the rest-frame 
near-infrared tend to be also brighter at 1.1 mm, and galaxies fainter than $\mchi=23$ do not produce detectable 
1.1 mm emission.  This suggests a correlation between stellar mass and SFR, but outliers to this 
correlation are also observed, suggesting strongly boosted star formation or extremely large 
extinction.  We also find tendencies that redder galaxies and galaxies at higher redshifts are brighter at 1.1 mm.
Our field contains $z\sim2.5$ H$\alpha$ emitters and a bright single-dish source.  
However, we do not find evidence of bias in our results caused by the bright source.
By combining the fluxes of sources detected by ALMA and fluxes of faint sources detected with stacking, 
we recover a 1.1 mm surface brightness of up to $20.3\pm1.2$ Jy deg$^{-2}$, 
comparable to the extragalactic background light measured by \emph{COBE}.
Based on the fractions of optically faint sources in our and previous ALMA studies
and the \emph{COBE} measurements, we find that approximately half of the cosmic star formation may be obscured by dust and
missed by deep optical surveys,    Much deeper and wider ALMA imaging is therefore needed to
better constrain the obscured cosmic star formation history.
\end{abstract}

\section{Introduction}

The extragalactic background light (EBL) is a measure of the radiative energy production from star formation
and black hole accretion throughout the history of the universe.  It is now known that the optical and far-infrared (FIR) 
portions of the EBL have comparable integrated strengths \citep[e.g.,][]{dole06}, implying that a large amount of the 
rest-frame UV radiation is absorbed by dust and reradiated in the FIR.  In order to understand the 
star formation history and accretion history fully, it is thus crucial to map the high-redshift dusty galaxies that give rise to the FIR EBL.  

Numerous deep imaging surveys have been carried out in the
millimeter and submillimeter (mm/submm) from the ground and in  the FIR from the space, to detect and study the 
FIR sources (see \citealp{casey14} and \citealp{lutz14} for recent reviews).  However, because of the effect of confusion of 
single-dish telescopes, the vast majority of the detected objects have infrared luminosity well above $10^{12}~L_\sun$,
corresponding to the bright end of the infrared luminosity functions. In the mm/submm, typically only 
10\%--40\% of the EBL is resolved into discrete bright sources by bolometer array cameras 
\citep[e.g.,][]{barger99,borys03,greve04,wang04,coppin06,weiss09,scott10,hatsukade11}.
In the FIR, \emph{Herschel} SPIRE surveys are only able to directly resolve $\sim15\%$ of the 200--500 $\mu$m 
EBL into bright sources \citep[e.g.,][]{oliver10}.  Imaging surveys in strong lensing clusters can nearly fully resolve 
the mm/submm EBL \citep[e.g.,][]{cowie02,smail02,knudsen08,chen13} and provide valuable insight into the nature of the 
faint sources \citep{chen14}.
However, the sample sizes for the lensed faint sources remain extremely small.

The advent of ALMA is transforming the studies of mm/submm sources.  ALMA does not only provide a powerful
means of following up the single-dish sources, but also serves as a survey machine. 
In particular, ALMA has the combination of high angular resolution and high sensitivity, the two key elements 
required to detect faint galaxies beyond the confusion limits of single-dish telescopes.   
In early ALMA  cycles, various small-scale continuum surveys have been conducted \citep[e.g.,][hereafter D16]{umehata15,dunlop16}.
However, because of the limited observing time, even these ALMA surveys did not 
reach the sensitivity required to fully resolve the EBL over large areas.
Sources detected in these ALMA surveys typically account for $\sim40\%$ of the EBL
\citep[e.g.,][hereafter H16; D16]{hatsukade16} and the majority of the dusty galaxies remain
undetected.  One way to break through the current sensitivity limit is,
instead of relying on contiguous ALMA mosaic survey, to exploit the archived data where the individual pointings
are sufficiently deep and to look for serendipitously detected faint objects (\citealp{hatsukade13,ono14,carniani15,fujimoto16}, hereafter F16; \citealp{oteo16}).  
Another way is to employ stacking analyses to obtain averaged mm/submm properties of high-redshift galaxies
\citep[e.g.,][D16]{decarli14,scoville14}.  Here we take the second approach and present stacking analyses of near-infrared (NIR) selected galaxies
in our ALMA 1.1 mm survey in the Subaru/XMM-Newton Deep Survey Field \citep[SXDF,][]{furusawa08}.

Our SXDF-ALMA survey covers an area of 2.0 arcmin$^2$ within the footprint of the Cosmic Assembly 
Near-IR Deep Extragalactic Legacy Survey \citep[CANDELS][]{grogin11,koekemoer11} in the SXDF. 
The extremely deep \emph{HST} WFC3 images provide large numbers of faint, NIR selected, high-redshift galaxies for stacking
analyses.    In Section 2, we describe the ALMA and multi-wavelength data. 
In Section 3, we describe the method of our stacking analyses and the results.  
In Section 4, we first examine whether a bright 1.1 mm single-dish source in our field biases our measurements.
We then estimate the contribution to the 1.1 mm EBL from the NIR and ALMA detected objects, compare our
results with previous studies, and discuss the implication.  All magnitudes are given in
the AB system, where $m_{AB} = 8.9 - 2.5\log(F)$ when flux $F$ is in unit of Jansky.
When we compare our 1.1 mm results with previous 1.3 mm and 870 $\mu$m results, we 
assume $F_{1.1\,\rm mm} = 1.65 \times F_{1.3\,\rm mm}$ (F16) and $F_{1.1\,\rm mm} = F_{870\,\mu \rm m}/2.5$ \citep{oteo16}.

\section{Data}

\subsection{SXDF-ALMA Survey}
\begin{figure}
\epsscale{1.2}
\plotone{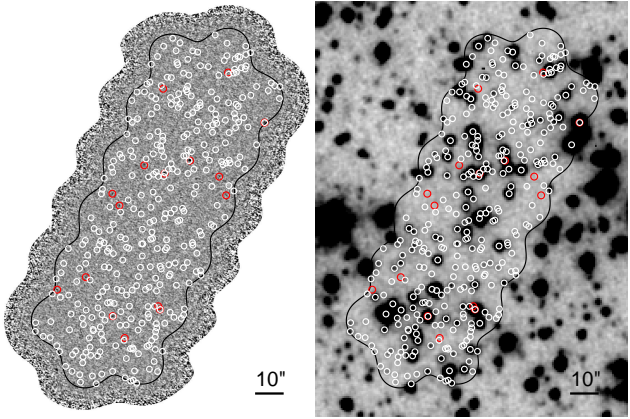}
\caption{Our ALMA 1.1 mm image (left) and \emph{Spitzer} IRAC 3.6 $\mu$m image (right) from the SEDS \citep{ashby13}.   
White circles show the positions of the 346 CANDELS WFC3 objects, and have $1\arcsec$ radii.
Red circles show the 16 $>4\sigma$ ALMA 1.1 mm sources.
The area enclosed by the contours have relative weights higher than the 75\% primary beam response.
\label{fig1}} 
\end{figure}

The observations of the SXDF-ALMA Survey (Program ID: 2012.1.00756.S; PI: K.\ Kohno)
and the data reduction will be described in K.\ Kohno et al.\ (in prep., see also, \citealp{kohno16}).  Here we provide
a brief summary and present the ALMA image in Fig.~\ref{fig1} (left). We conducted Band 6 (1.1 mm, or 274 GHz) continuum imaging in the SXDF in ALMA Cycle 1, 
with a 19-pointing mosaic, a total bandwidth of 7.5 GHz, and a total observing time of 3.6 hr.  
The field is selected to cover a bright AzTEC 1.1 mm source and 12 $z\sim2.5$ H$\alpha$-selected 
star-forming galaxies \citep[e.g.,][]{tadaki15}.   The calibration and imaging are performed with the Common Astronomy Software Application 
package \citep{mcmullin07}.  The visibility data were naturally weighted to produce a CLEANed map with a synthesized beam of
$0\farcs53 \times 0\farcs41$ (PA = $64\arcdeg$).  In this work, we only consider the deep region where the effective coverage is 
greater than 75\% of the peak primary beam response, indicated by the contours in Fig.~\ref{fig1}.  
This excludes a bright objects near the map edge (SXDF-ALMA 3 in \citealp{yamaguchi16}).
The area in this region is 1.58 arcmin$^2$ and the typical rms noise is 62 $\mu$Jy beam$^{-1}$.
There are 16 sources detected in this area at $>4~\sigma$, and eight sources at $>4.5~\sigma$.  
Up to 1/3 of the $>4~\sigma$ sources could be spurious, based on the number of negative peaks (H16),
and the number of spurious sources decreases to zero at $>4.7~\sigma$.

\subsection{Optical and NIR Data}
Our stacking analyses are based on the WFC3 detected objects in the CANDELS catalog of \citet{galametz13}.
This catalog includes \emph{Spitzer} IRAC fluxes of the WFC3 objects, extracted from the images of the \emph{Spitzer} Extended Deep Survey (SEDS, \citealp{ashby13}) at
the positions of the WFC3 sources (see \citealt{galametz13} et al. for details).
In Section~\ref{sec_flux_stack} we will show that the IRAC fluxes trace faint 1.1 mm emission better than the WFC3 fluxes, and therefore the majority of
our analyses will be based on an additional 3.6 $\mu$m selection in the CANDELS catalog.  The 5-$\sigma$ limiting magnitudes in our ALMA area for the $F160W$ and 
IRAC 3.6 $\mu$m bands are approximately 27.0 and 25.3, respectively.   The IRAC fluxes were extracted at the WFC3 positions, so objects can be assigned IRAC fluxes
much fainter than the nominal detection limits but with low S/N.  In this work, we consider objects fainter than $\mchi=26.0$ undetected at 3.6 $\mu$m.
Our visual inspection of the IRAC images does not find any
IRAC objects undetected by WFC3 (Fig.~\ref{fig1} (right)).  Therefore, our 3.6 $\mu$m selection from the WFC3 pre-selected catalog is not biased against any red objects.

There are 346 CANDELS objects in the 1.58 arcmin$^2$ area of our ALMA image, among which 151 of them have $>5\sigma$ fluxes at 3.6 $\mu$m and
197 of them have $\mchi < 26.0$.  We adopt a $0\farcs5$ search radius (approximately the beam FWHM) for our counterpart identification in this work.
Among the eight $>4.5~\sigma$ ALMA sources, four have CANDELS counterparts.  They are discussed in more details in \citet{yamaguchi16}.
The other four $>4.5~\sigma$ ALMA sources have 1.1 mm fluxes of 0.22--0.32 mJy.
These eight sources are not included in most of our stacking analyses but they are included in our analyses of the EBL contribution (Sec~\ref{sec42}).  In the 4--4.5~$\sigma$ range, there 
are eight ALMA sources, and only one has a CANDELS counterpart.  
This implies either a high spurious rate at 4--4.5~$\sigma$, or an extremely dusty population whose NIR light is extinguished.
Our analyses in H16 indicate a spurious fraction of $\lesssim40\%$.  This leaves roughly four real sources that are highly
obscured in the optical and rest-frame NIR.

We supplement the CANDELS NIR catalog with our own photometric redshifts.
The optical to 4.5 $\mu$m data used for our photometric redshifts are similar to those included in the Galametz et al.\ catalog,
except that we also use the \emph{Galex} NUV data and our own CFHT MegaCam $U$-band data obtained in a 
multi-year $U$-band imaging campaign for the SXDF (W.-H.\ Wang, in prep).  The photometric redshifts have a
very good overall accuracy of $\Delta z/(1+z)=0.026$ on $K_s<25.0$ objects.  Among the 346 CANDELS objects in
the ALMA area, 147 (42\%) have reliable photometric redshifts whose $\chi^2$ are sufficiently low. 
The photometric redshift completeness increases substantially to 89\% at $\mchi<25.5$.
In the bright end of $\mchi < 23$, 46 out of 53 sources have photometric redshifts after we supplement three photometric redshifts 
from \citep{caputi11}.  Nearly all of them are at $z=0.5$--3.0.

\section{Stacking Analyses and Results}\label{section_stacking}

\subsection{Method}

Our stacking analyses method is very similar to those developed by \citet{wang12} and \citet{to14} for Very Large Array images.
To avoid biases caused by the small number of brighter objects, we exclude the four NIR objects that are detected at $>4.5$~$\sigma$ by ALMA.
We mask them and also the four $>4.5\sigma$ ALMA sources that do not have NIR counterparts in the image. 
This way, the $>4.5\sigma$ do not bias the measured fluxes and noise in out stacking analyses, and
our stacking results are more representative to the faint 1.1 mm population.
We averaged the 1.1 mm images centered at the NIR
objects, or averaged the 1.1 mm fluxes measured at the positions of the NIR objects.  Averaging the images and fluxes should give
identical results, but the former allows us to examine the image and to examine the averaged size of the objects.
In the stacked image, the mean background as well as background rms can be directly measured.
In the flux stacking, we estimated the uncertainties with a Monte Carlo method.  We placed random apertures in the image
and measured their mean flux.  The number of random apertures was identical to the number of NIR sources.  
There is a finite probability for random apertures to be located near bright objects.  If their fluxes exceed 4.5~$\sigma$,
they are considered as ``detections'' and removed from the random sample, just like what we do on the NIR objects.  
When we estimate the total contribution to the 1.1 mm EBL, we account for these detected sources separately.
We repeated this for $10^5$ times and calculated the mean and standard deviation
of the $10^5$ mean fluxes.  The mean is considered as a background value and subtracted from the measured mean flux
of the NIR selected sources. This statistically removes the effect of faint confusing sources and uncleaned sidelobes.
The standard deviation is considered as the uncertainty in the mean flux of the NIR selected sources.
Finally, to test whether our stacking results may be biased, we injected artificial point sources to the image with random positions.
Each time we injected tens to $<200$ sources, measured their stacked fluxes, and repeated $10^4$ times.
We did not find any systematic bias in the average of the $10^4$ mean fluxes for input fluxes ranging from 0.005 mJy to 0.2 mJy.

\subsection{Results of Image Stacking}\label{sec_image_stack}

\begin{figure}
\epsscale{1.2}
\plotone{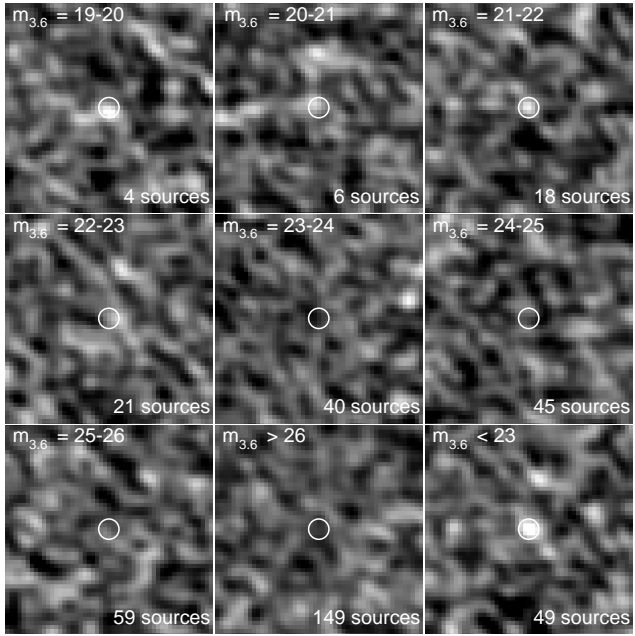}
\caption{Stacked 1.1 mm images centered at CANDELS WFC3 sources (small circles, $D=0\farcs5$) in various $\mchi$ bins. 
Each panel is $5\arcsec$ on a side.  The brightness scale of each panel is from $-2~\sigma$ to $+4~\sigma$.  
See Table~\ref{tab1} for $\sigma$ and stacked fluxes.  Note that there are 11 $>4\sigma$ ALMA-detected sources without
WFC3 counterparts, and therefore not included in any of these panels.
\label{fig2}} 
\end{figure}

We present stacked ALMA 1.1 mm images at the CANDELS 3.6 $\mu$m sources in Fig.~\ref{fig2}.  If we simply stack all sources, 
we do not reach a $>3~\sigma$ detection.  This indicates that most of the faint 3.6 $\mu$m sources do not exhibit
strong dust emission.  On the other hand, once we split the samples according to their 3.6 $\mu$m magnitudes and stack,
we see a clear trend that brighter NIR sources are also brighter 1.1 mm sources, on average.  This is generally true in all the NIR bands
from F125W to 4.5 $\mu$m, and the tendency is stronger in the two IRAC bands.  We will describe this in more details in the next subsection.
From Fig.~\ref{fig2}, it can also be seen that the apparent 1.1 mm positions do not always exactly match the NIR positions (circles in the figure).
However, the offsets are all smaller than our $\sim0\farcs5$ ALMA beam in the top four $\Delta\mchi=1$ bins.   This can be explained
with the relatively low S/N.   If we stack all sources with $\mchi<23$ (lower-right panel of Fig.~\ref{fig2}, which gives the maximum S/N, the offset disappears.

Our stacked 1.1 mm flux for the 28 $\mchi<22$ sources is $\langle F_{\rm1.1\,mm} \rangle = 0.067\pm0.013$ mJy.  In \citet{decarli14}, the stacked 344 GHz for 85 $K<22$ 
and $F_{344\,GHz} < 1.2$ mJy sources is $0.20\pm0.06$ mJy.  This is fully consistent with ours once our 1.1 mm stacked flux is scaled by $2.5\times$
\citep{oteo16}, given that in this magnitude range the typical $K-\mchi$ color is $\sim1.0$.  This is also comparable to the faintest stacked 350 GHz
flux reached in \citep{scoville14}.

We measured the size of detected 1.1 mm emission in the stacked images from the $\mchi < 23$ bin 
($\langle F_{\rm1.1\,mm} \rangle = 0.050\pm0.010$) by 
fitting a 2D Gaussian distribution.  The fitted Gaussian has a FWHM of $0\farcs76 \times 0\farcs43$, slightly larger than
the $0\farcs53 \times 0\farcs41$ synthesized beam.  This suggests small source sizes, probably only marginally resolved by the ALMA beam.
This is consistent with higher angular resolution ALMA studies of  brighter sources (\citealp{simpson15,ikarashi15}) and also
the stacking analyses in D16.  However, given the low S/N and the apparent offsets between the NIR positions and the stacked
1.1 mm positions in Fig.~\ref{fig2}, we do not think we can place a meaningful constraint on the source size. 
We only use this result to guide our selection of flux aperture size in our subsequent analyses.

\subsection{Results of Flux Stacking}\label{sec_flux_stack}

We move on to measure 1.1 mm fluxes at the positions of the CANDELS sources and stack the fluxes.
Based on the above-measured source size, we adopted a relatively small flux aperture, $D=0\farcs5$ (circles in Fig.~\ref{fig2}).
The aperture correction is derived based on the dirty beam assuming unresolved sources.
Such a small aperture gives higher S/N while still enclosing most of the fluxes from unresolved and slightly resolved objects.
It does not require different aperture corrections for CLEANed
and unCLEANed sources, since the clean and dirty beams only become substantially different
at distances greater than a FWHM.  The major results are summarized in Table~\ref{tab1}.

\begin{figure}
\epsscale{1.05}
\plotone{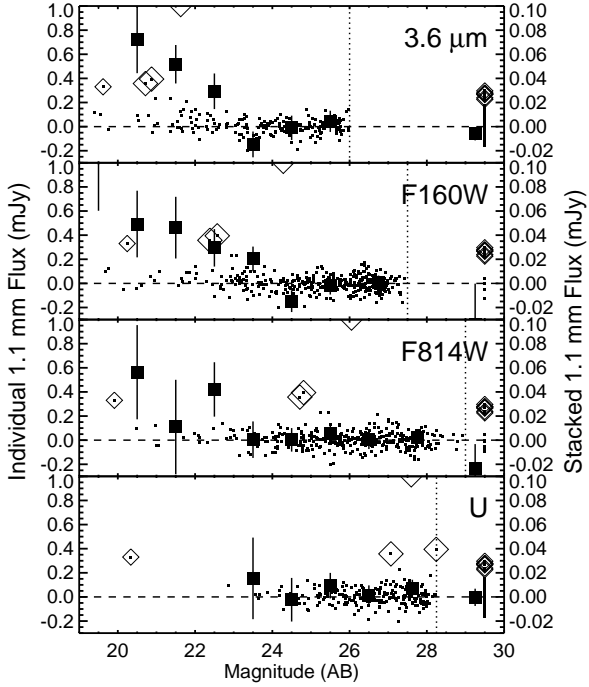}
\caption{1.1 mm flux vs. $U$, $F814W$, $F160W$, and 3.6 $\mu$m magnitudes.  The small symbols and \emph{left}-$y$-axis
represent fluxes of individual objects.  The large filled squares with error bars and the \emph{right}-$y$-axis represent stacked fluxes
in  $\Delta m=1$ bins.  Large diamonds are individually detected $>4.5~\sigma$ sources and are not included in the stacking analyses. 
Small diamonds are 4--4.5~$\sigma$ sources. Vertical dotted lines show the detection limits at $U$ to 3.6 $\mu$m. Objects not detected in these
bands are shown at $m = 29.5$ and their stacked fluxes are shown at $m = 29.25$.
The stacked results of the brightest $F160W$ and 3.6 $\mu$m bins are outside the range of the plots (see Fig.~\ref{fig4} 
for the 3.6 $\mu$m one).
\label{fig3}} 
\end{figure}

In Fig.~\ref{fig3} we present 1.1 mm fluxes of individual galaxies and stacked 1.1 mm fluxes vs.\ magnitudes
at $U$, $F814W$, $F160W$, and 3.6 $\mu$m bands.  When the 1.1 mm fluxes are ordered according to the $U$-band magnitudes, we
cannot obtain detections with stacking analyses.  Some of the brighter 1.1 mm sources are even not detected at $U$ 
and therefore not included in the $U$-band stacking analyses.  This shows that the unobscured star formation
(traced by rest-frame UV) does not strongly correlate with obscured star formation (traced by dust emission at 1.1 mm).
On the other hand, as we move to longer wavebands, there is a tendency that objects more luminous in the 
rest-frame NIR are brighter 1.1 mm sources.  This trend is the strongest at 3.6 $\mu$m.  This is consistent with 
the results in Fig.~\ref{fig2}.  The strong correlation
between 3.6 $\mu$m magnitudes and stacked 1.1 mm fluxes can be explain by a correlation between stellar mass
and obscured star formation.  On the other hand, ongoing star formation can also
boost the rest-frame NIR luminosity of galaxies and lead to a correlation between 3.6 $\mu$m and 1.1 mm.  
However, this effect should be even stronger in the optical bands ($U$ and
$F814W$ in Fig.~\ref{fig3}) as young stellar populations are blue, and this is not observed.  
Stronger dust extinction in the bright end may hinder this effect in the 
optical, but the lack of correlation between 3.6 $\mu$m magnitudes and galaxy colors in the optical (which should
be more strongly affected by extinction) in our samples does not support this scenario.  Therefore, a correlation between stellar mass and star formation 
is a more plausible and natural explanation
for the correlation between 3.6 $\mu$m and 1.1 mm.  This is consistent with the recent results in D16.

\begin{figure}
\epsscale{1.0}
\plotone{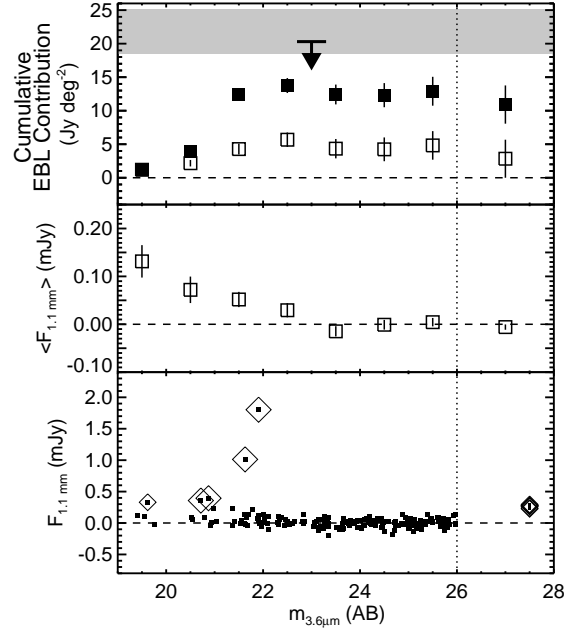}
\caption{1.1 mm fluxes and EBL contributions vs.\ 3.6 $\mu$m magnitudes.  Bottom panel: 1.1 mm fluxes of individual
objects.  The four $>4.5~\sigma$ detections are enclosed by large open diamonds and they are not included in 
the subsequent stacking analyses. The 4--4.5~$\sigma$ ALMA-detected objects are enclosed by small open diamonds,
and are placed at $\mchi = 27.5$ if they do not have WFC3 counterparts 
(11 sources).  The vertical dotted line shows the 3.6 $\mu$m detection limit.
Middle panel: stacked flux and flux error in each $\Delta \mchi = 1$ bin.  
Top panel: cumulative contribution to the EBL for from the NIR selected objects.
The open squares show the contribution from $<4.5~\sigma$ fluxes.
The solid squares show the contribution from all NIR selected objects (i.e., including the four open diamonds in 
the bottom panel).  We do not show the individual 1.1 mm fluxes of WFC3 objects undetected at 3.6 
$\mu$m (149 objects), and place their stacked flux and EBL contribution at $\mchi = 27$.
The horizontal bar includes the contribution of all $>4.0~\sigma$ ALMA sources 
and this is an upper limit (see Section~\ref{sec42}).
The shaded band shows the range of the 1.1 mm EBL measured by \emph{COBE}.
\label{fig4}} 
\end{figure}

The above observed correlation between $\mchi$ and stacked 1.1 mm flux is not universal and is probably only applicable to
faint mm/submm sources, on average.  At least two of the $>4.5\sigma$ detections (two brightest open diamonds in Fig.~\ref{fig4}) have 1.1 mm 
fluxes significantly above the correlation.  There also exist additional four $>4.5 \sigma$ sources that do not have counterparts in the CANDELS
catalog and therefore not included in Fig.~\ref{fig4}.  They should have $\mchi > 26$ and thus are also highly above the correlation.  
Either dramatically boosted SFR or extremely large extinction (or both) can explain these objects.  Either of the possibilities should not
be surprising for bright mm/submm sources.

In the $\mchi=20$--22 range, (24 sources) the stacked flux of $0.057\pm0.014$ mJy corresponds to an infrared luminosity of 
$\sim9\times10^{10}~L_\sun$ for $z=2$ if we adopt the luminosity-dependent dust spectral energy distribution (SED) in
\citet{chary01}, or $\sim3\times$ larger if assume an Arp 220 SED.
The star formation rate (SFR) derived with the \citet{chary01} SED and a \citet{kennicutt98} conversion is $\sim 15~M_\sun$ yr$^{-1}$. 
The values of infrared luminosity and SFR probed by the $\mchi=22$--23 stacking samples are $2\times$ smaller, with a lower
2 $\sigma$ significance.  
The above stacked SFRs should be diluted by an unknown fraction of quiescent galaxies, otherwise the face values 
are similar to the UV SFR of faint Lyman-break galaxies.

\begin{deluxetable}{cccccc}
\tablecaption{Results of Stacking Analyses\label{tab1}}
\tablehead{\colhead{NIR Sample}  & \colhead{$N$}  & \colhead{$\langle F_{\rm1.1\,mm} \rangle$ (mJy)} & \colhead{EBL (Jy deg$^{-2})$}}
\startdata
$\mchi =19$--20	& 4		& $0.131\pm0.034$		& $1.19\pm0.31$ \\
$\mchi =20$--21	& 6		& $0.072\pm0.028$		& $0.98\pm0.38$ \\
$\mchi = 21$--22 	& 18		& $0.052\pm0.016$		& $2.12\pm0.65$ \\
$\mchi = 22$--23	& 21		& $0.029\pm0.015$		& $1.39\pm0.70$ \\
$\mchi = 23$--24	& 40		& $-0.015\pm0.011$		& $-1.34\pm0.97$ \\
$\mchi = 24$--25	& 45		& $-0.001\pm0.010$		& $-0.10\pm1.03$ \\
$\mchi = 25$--26	& 59		& $0.004\pm0.009$		& $0.59\pm1.18$ \\
$\mchi > 26$		& 149	& $-0.006\pm0.006$		& $-1.99\pm3.26$ \\
$\mchi < 23$		& 49		& $0.051\pm0.010$		& $5.67\pm1.07$ \\
\hline
$K_s-\mchii<0$	& 18		& $0.044\pm0.016$		& $1.79\pm0.65$ \\
$K_s-\mchii=0$--1	& 23		& $0.029\pm0.014$		& $1.51\pm0.73$ \\
$K_s-\mchii>1$	& 8		& $0.131\pm0.024$		& $2.37\pm0.43$ \\
\hline
$z = 0$--2			& 33		& $0.040\pm0.012$		& $2.70\pm0.80$\\
$z = 2$--4			& 13		& $0.097\pm0.019$		& $2.86\pm0.56$
\enddata
\tablecomments{All the values listed here do not include the four $>4.5\sigma$ ALMA detected sources.
The color and redshift subsamples only include $\mchi<23$ objects.}
\end{deluxetable}

We now focus on stacking analyses based on the 3.6 $\mu$m magnitude as it provides the strongest 1.1 mm signal.
Fig.~\ref{fig4} shows the measured 1.1 mm fluxes vs. 3.6 $\mu$m magnitudes.
Both Fig.~\ref{fig3} and Fig.~\ref{fig4} show that sources fainter than $\mchi=23$ do not produce detectable 1.1 mm emission
after stacking (except for those ALMA detected sources without WFC3 counterparts).  
This does not change if we adopt photometric apertures as large as $2\arcsec$, implying that the non-detections
are not caused by random positional offsets between the NIR and the 1.1mm emission.

We estimate the contribution to the 1.1 mm EBL from the stacked objects by dividing their integrated
flux with survey area (top panel of Fig.~\ref{fig4}). At $\mchi<23$, the cumulative contributions are 
$5.67\pm1.07$ Jy deg$^{-2}$ from  sources with $<4.5~\sigma$ ALMA fluxes, and $13.75\pm1.12$ Jy deg$^{-2}$
from all  sources. The 1.1 mm EBL measured by the \emph{COBE} FIRAS experiments is  
18.5 Jy deg$^{-2}$ \citep{puget96} or 25.1 Jy deg$^{-2}$ \citep{fixsen98}.  
This range of 18.5--25.1 Jy deg$^{-2}$ probably reflects
the systematic uncertainty in the measurements, and is the shaded area in the top panel of Fig.~\ref{fig4}.
Our recovered EBL from all $\mchi<23$ sources correspond to 60\% to 80\% of the COBE values.
We will further discuss this in Section~\ref{sec4}.

\begin{figure*}
\epsscale{1.05}
\plotone{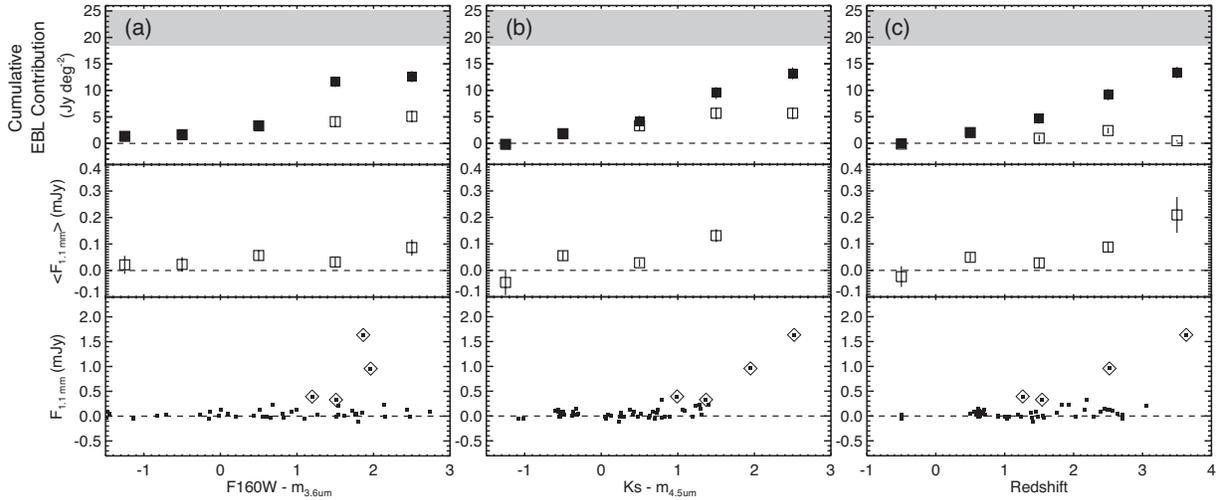}
\caption{Same as Fig.~\ref{fig4}, but for 1.1 mm fluxes and EBL contributions vs.\  $F160W-\mchi$ color (a), vs.\ $K_s-\mchii$ color (b), and vs.\ redshift (c), 
for objects with $\mchi<23$. ALMA detected objects (open diamonds) are not included in the stacked fluxes and
EBL contributions (open squares).  Objects with no reliable photometric redshifts are assigned $z=-0.5$.\label{fig5}} 
\end{figure*}

We can use stacking analyses to examine the 1.1 mm flux as functions of NIR color on the 49 sources with $\mchi<23$. 
In previous studies, it is shown
that bright mm/submm sources tend to be redder in the NIR and IRAC wavebands 
(e.g., \citealp{chen16}; and likely vice versa, e.g., \citealp{wang12,caputi14}).
Fig.~\ref{fig5} (a) and (b) shows that this is also the case for faint sources (also see Table~\ref{tab1}).  The four $>4.5~\sigma$ ALMA sources with
NIR counterparts all have moderately red $F160W-\mchi$ colors and extremely red $K_s-\mchii$ colors. The stacked fluxes also suggest a tendency
that redder sources are brighter at 1.1 mm, although this is primarily driven by the sources in
the  $F160W-\mchi =2$--3 bin and $K_s-\mchii=1$--2 bin.  

Finally, we present the redshift dependence of the stacked 1.1 mm flux of the 46 sources with redshifts and $\mchi<23$
in Fig.~\ref{fig5}(c) (also see Table~\ref{tab1}).  The diagram remarkably resembles that in Fig.~\ref{fig5}(b), because the
$K_s-\mchii$ color almost monotonically increases with redshift (see, e.g., Fig.~1 in \citealp{wang12}) at
$z<3$.  Similarly, the trend is primarily driven by the few sources at higher redshifts, and the stacked
fluxes of fainter objects are noisy.

\section{Discussion}\label{sec4}

\subsection{The Bright AzTEC Source}\label{sec41}
Our ALMA field is not an unbiased blank field, as it was chosen to include a bright, 3.5 mJy AzTEC source (S.\ Ikarashi et al., in preparation).  
This source splits into two sources under ALMA's resolution (the two brightest sources in the bottom panels of Fig.~\ref{fig4}/\ref{fig5}). 
The total flux of them measured by \citet{yamaguchi16} is $3.4\pm0.18$ mJy, consistent with the AzTEC flux.
They both show excess of emission in the 2315 nm narrow-band filter \citep{tadaki15},
suggesting an H$\alpha$ redshift of $z=2.53$.  This redshift is consistent with their photometric redshifts \citep{yamaguchi16} given the large
photometric redshift errors, but needs to be confirmed with spectroscopy.   We expect to enclose $\sim0.1$ such objects in our 
ALMA field, based on its AzTEC flux and the blank field AzTEC counts in \citet{scott12}, if our field is randomly placed.
From this point of view, it is a rare object and the over-density associated with it may bias our results.  On the other hand, if we look at them individually
based on their ALMA fluxes, we expect to find $\sim0.3$ and $\sim0.8$ sources in our ALMA field based on the AzTEC counts.  
Then the existence of such sources in our survey may not be too surprising.

\begin{figure}
\plotone{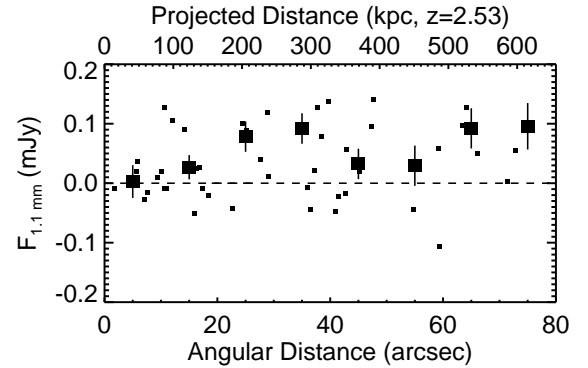}
\caption{1.1 mm fluxes of $\mchi<23$ galaxies vs.\ projected distance from the bright AzTEC source (measured from the center of the two bright ALMA sources).
Small symbols are individual fluxes, and solid squares with error bars are stacked fluxes.  The \emph{top}-$x$-axis is the projected physical distance
at $z=2.53$, the H$\alpha$ redshift of the two bright ALMA sources.
\label{fig6}} 
\end{figure}

Nevertheless, it is possible that our stacking results are biased, especially in the area around the 3.5 mJy AzTEC source.
This concern arises from the result that bright mm/submm sources are strongly clustered and reside in massive dark matter halos\citep[e.g.,][]{hickox12,chen16}.  
We can test this with our data.  In the ALMA studies of color-selected sources in \citet{decarli14}, the authors found tentative evidence that galaxies within 200 kpc 
from bright 870 $\mu$m sources tend to be also brighter at 870 $\mu$m.  
We do not detect such a trend on the 3.6 $\mu$m sources in our ALMA field.  In Fig.~\ref{fig6}, we show 1.1 mm fluxes vs.\ projected distance from the bright AzTEC source
(measured from the center of the two bright ALMA sources) on $\mchi<23$ objects.  We see no evidence of elevated 1.1 mm flux near the bright source
within the scales probed by our ALMA imaging.  This does not change even if we include fainter 3.6 $\mu$m objects.
The difference between the results here and those in \citet{decarli14} could be caused by sample sizes 
(one bright AzTEC source here vs.\ $>100$ LABOCA sources in Decarli et al.).

\begin{figure}
\plotone{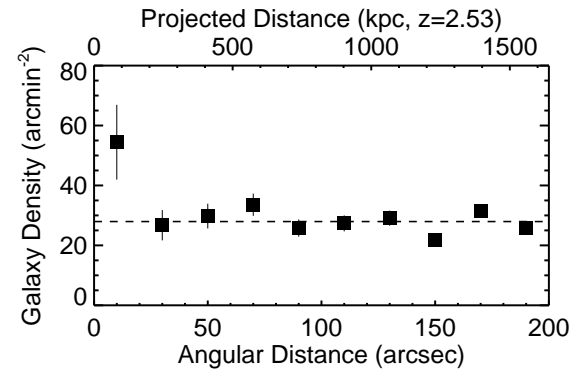}
\caption{Similar to Fig.~\ref{fig6} but for surface density of $\mchi<23$ galaxies.
The error bars are Poisson errors in the galaxy counts.
\label{fig7}} 
\end{figure}

We also investigate the number density of $\mchi<23$ galaxies as a function of projected distance from the bright AzTEC source.
The result is shown in Fig.~\ref{fig7}.  There is a 2~$\sigma$ over-density within $20\arcsec$ around the bright AzTEC source.
This over-density can be observed until $\mchi<24$, and then the area becomes under-dense in $24<\mchi<26$.
The combined density distribution becomes more or less flat for all galaxies with $\mchi<26$.  Moreover, using the photometric redshifts
($\sim30\%$ complete at $\mchi<26$), we found that galaxies with photometric redshifts of $<2.0$ contribute 
$21\pm10$ arcmin$^{-2}$ to the surface density above the large-scale average at $r<20\arcsec$. 
The over-density at $r<20\arcsec$ in Fig.~\ref{fig7} is thus primarily driven by galaxies unrelated to the AzTEC source,
and is probably a result of small number statistics.
We conclude that there is no solid evidence of over-density around the bright AzTEC source.  Even if the over-density in Fig.~\ref{fig7} is real, 
after being multiplied by the low stacked flux shown in Fig.~\ref{fig6}, it does not alter our results of mean 1.1 mm fluxes of galaxies or the 
EBL contribution.  In our discussion below, we do not make special treatments to galaxies near the AzTEC source.  However, we consider
results derived with and without the two bright ALMA sources that compose of the AzTEC source.

\subsection{Resolved EBL}\label{sec42}
\begin{figure}
\epsscale{1.1}
\plotone{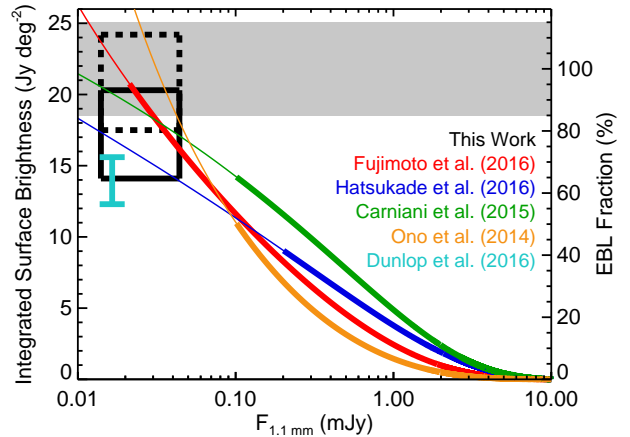}
\caption{Resolved 1.1 mm EBL.  Our stacking analyses plus ALMA detected sources in the SXDF is shown
with the solid box. The top and bottom ends of the box correspond to the upper and lower limits discussed
in Section~\ref{sec42}.  The horizontal position and width are the stacked flux and flux errors, respectively, 
of the NIR sources in the $\mchi=22$--23 bin. 
The dashed box includes the contribution from a hypothesized optically faint population discussed in Section~\ref{sec43}.
Color curves show the cumulative EBL contributions derived from Schechter fits to
number counts in recent ALMA studies.  The thick portions of the curves indicate the flux range probed by 
these studies and the thin portions are the extrapolations.  The cyan error bar is the 1.3 mm result from D16, scaled 
by $1.65\times$ to 1.1 mm (F16).
The shaded band shows the range of the 1.1 mm EBL measured by \emph{COBE},
and the middle point is labeled as 100\% along the right-$y$ axis.
\label{fig8}} 
\end{figure}

A key question we would like to address with our SXDF-ALMA survey is how much of the 1.1 mm EBL can be 
directly detected (see H16) and recovered with stacking analyses.  An EBL contribution of $13.75\pm1.12$ Jy deg$^{-2}$ 
is recovered if we include all NIR sources, which is shown by the solid squares in the top panel of Fig.~\ref{fig4}.  
We can further include ALMA detected sources 
without NIR counterparts (i.e., not included in the stacking analyses).  The results are $16.1 \pm 1.2 $ Jy deg$^{-2}$
and $20.3\pm1.2$ Jy deg$^{-2}$, respectively, for including $>4.5\sigma$ and $>4\sigma$ sources.
The latter is the downward pointing arrow in the top panel of Fig.~\ref{fig4} and the upper end of the solid box in Fig.~\ref{fig8}.
All these values are upper limits for the following reasons.  First, our ALMA field is chosen to include a bright AzTEC 
source.  Based on the blank-field counts in \citet{scott12}, we expect $<1$ such sources in our ALMA field, no matter
for a single, bright AzTEC source or for the two ALMA sources that it splits into.
Second, between 4--4.5~$\sigma$, the combined effect of flux boosting ($\sim$15\%--20\%), spurious sources ($\lesssim40\%$), 
and completeness ($\sim70\%$) may overestimate the contribution in the 4--4.5~$\sigma$ interval (H16).

A more direct way to account for all the above effects is to involve our number counts in H16, which took into account
the flux boosting, spurious fraction, and completeness.
The bright end of the counts in H16 is supplemented by single-dish counts and not entirely based on the SXDF-ALMA data.  
The H16 counts fitted with a Schechter function integrated to 0.2 mJy yield an EBL contribution of 9.2 Jy deg$^{-2}$, which is 
represented by the thick blue curve in Fig.~\ref{fig8}.\  After removing sources already accounted for in H16,
our stacked EBL contribution from NIR sources without ALMA detections is 4.9 Jy deg$^{-2}$.
Therefore, the combined EBL contribution from bright and faint sources is 14.1 Jy deg$^{-2}$.
This is the lower end of the solid box in Fig.~\ref{fig8}.  This is comparable with the result in D16,
who also employ stacking analyses to supplement the directly detected fluxes.

The solid box in Fig.~\ref{fig8} represents the range of resolved EBL probed by our stacking of faint objects, plus two different
treatments of bright objects (total detected flux in our field or adopting the wide-field bright-end counts).  
It is broadly consistent with that in F16, who conducted the most thorough search by far for serendipitously 
detected faint continuum sources in deep ALMA 1.2 mm pointings in the archive.  The counts in F16 at a $>0.02$ mJy level 
lead to an EBL contribution of $22.9^{+6.7}_{-5.6}$ Jy deg$^{-2}$, while we recover up to 20.3 Jy deg$^{-2}$
of EBL at 1.1 mm with stacking analyses on a faint population with $\langle F_{\rm 1.1\,mm} \rangle \sim0.03$ mJy.
The caveat here is that the F16 sample includes 1.2 mm sources that are optically faint, while our stacked signal 
comes from relatively bright NIR sources ($\mchi<23$).  This leads to our next discussion topic.

\subsection{Optically Faint Sources}\label{sec43}

Once we detect the majority of galaxies that give rise to the 
mm/submm EBL, we would like to ask what kinds of galaxies they are.  Additional to color and redshift distributions of
these galaxies, of particular interest is the fraction of optically faint galaxies, i.e., galaxies missed by 
deep optical/NIR surveys.  This  tells us whether the star formation history constructed from optically
selected galaxies is representative,
or needs significant revision.  We can gain insight on this by comparing our results with the results of F16 and with the \emph{COBE} EBL values.

The EBL resolved by our ALMA imaging and stacking analyses is still lower than the EBL measured by \emph{COBE}.
This suggests a considerable fraction (anywhere between 0 and 44\%) of EBL arising from sources fainter than the CANDELS detection limit.
Such extremely optically faint sources have been found in previous surveys \citep[e.g.,][]{wang09}, as well as our SXDF-ALMA survey
(SXDF-ALMA3 
in \citealp{yamaguchi16}).  They are also hinted by the large number of ALMA detected sources without
CANDELS counterparts (four out of eight for $F_{\rm 1.1\,mm}>4.5~\sigma$).  
The optically faint fraction (50\%) is similar to that in F16 (41\%).   

Because of the above, we hypothesize an optically faint population that is not picked up by our NIR selection.  
We further assume that this population accounts for $\sim41\%$ of the EBL 
from the faint end, based the optically faint fraction in F16, for its larger sample size and higher ALMA sensitivity.  
This means a completeness correction of $1/(1-0.41)$ to our NIR stacked EBL.  
Once we do so, the solid box in Fig.~\ref{fig8} becomes
the dashed box, corresponding to 17.5 to 24.2 Jy deg$^{-2}$.
This is comparable to the range allowed by the \emph{COBE} measurements and the range probed/extrapolated by 
previous number counts.
We therefore conclude that an optically faint fraction in the ballpark
of 50\%--60\% is consistent with existing data for both the bright and faint ends of the 1.1/1.2 mm population.  If this is the case, then 
optical studies can only account for some 50\% of high-redshift star-forming galaxies.  

The above studies demonstrate that with existing ALMA data, we just barely can scrape the surface 
of the issues raised by the resolved EBL and extremely dusty galaxies.  Future ALMA deep imaging 
will be able to put better constraints on the optically faint fraction as functions of mm/submm fluxes.
Ultimately, the accuracy in ALMA determinations of the EBL contribution 
from discrete sources may even exceed that in the \emph{COBE} measurements.
These will further transform our understanding of the dusty side of the galaxy evolution.

\acknowledgments
We thank Seiji Fujimoto for the useful discussion and the referee for the thorough report.
WHW was a resident astronomer at the CFHT when the majority of the analyses were carried out.
WHW is supported by the Ministry of Science and Technology of Taiwan (102-2119-M-001-007-MY3).
BH, YT, and YM are supported by the Japan Society for Promotion of Science (JSPS) KAKENHI (Nos. 15K17616, 25103503, 15H02073, 20647268). KK is supported by the ALMA Japan Research Grant of NAOJ Chile Observatory, NAOJ-ALMA-0049. KC and SI acknowledge the support of the Netherlands Organisation for Scientific Research (NWO) through the Top Grant Project 614.001.403. JSD acknowledges the support of the European Research Council via an Advanced Grant. ML is financially supported by a Research Fellowship from JSPS for Young Scientists. HU acknowledges the support from Grant-in-Aid for JSPS Fellows, 26.11481. This paper makes use of the following ALMA data: ADS/JAO.ALMA\#2012.1.00756.S. ALMA is a partnership of ESO (representing its member states), NSF (USA) and NINS (Japan), together with NRC (Canada), NSC and ASIAA (Taiwan), and KASI (Republic of Korea), in cooperation with the Republic of Chile. The Joint ALMA Observatory is operated by ESO, AUI/NRAO and NAOJ.
This paper is partially based on observations obtained with MegaPrime/MegaCam, a joint project of CFHT and CEA/DAPNIA, at the CFHT which is operated by the National Research Council (NRC) of Canada, the Institut National des Sciences de l'Univers of the Centre National de la Recherche Scientifique of France, and the University of Hawaii.

\end{document}